# A Food Recommender System in Academic Environments Based on Machine Learning Models

**Abolfazl Ajami**

MSc Information Technology, Tarbiat Modares University, E-mail: A.ajami@modares.ac.ir

**Dr. Babak Teimourpour** ⁎

Assistant Professor at the Faculty of Industrial and Systems Engineering Tarbiat Modares University, E-mail:

b.teimourpour@modares.ac.ir

## Abstract

**Background:** People's health depends on the use of proper diet as an important factor. Today, with the increasing mechanization of people's lives, proper eating habits and behaviors are neglected. On the other hand, food recommendations in the field of health have also tried to deal with this issue. But with the introduction of Western nutrition style and the advancement of Western chemical medicine, many issues have emerged in the field of disease treatment and nutrition. Recent advances in technology and the use of artificial intelligence methods in information systems have led to the creation of recommender systems in order to improve people's health.

**Methods:** A hybrid recommender system including, collaborative filtering, content-based, and knowledge-based models was used. Machine learning models such as Decision Tree, $k$-Nearest Neighbors (kNN), AdaBoost, and Bagging were investigated in the field of food recommender systems on 2519 students in the nutrition management system of a university. Student information including profile information for basal metabolic rate, student reservation records, and selected diet type are received online. Among the 15 features collected and after consulting nutrition experts, the most effective features are selected through feature engineering. Using machine learning models based on energy indicators and food selection history by students, food from the university menu is recommended to students.

**Results:** The AdaBoost model has the highest performance in terms of accuracy with a rate of 73.70%.

**Conclusion:** Considering the importance of diet in people's health, recommender systems are effective in obtaining useful information among a huge amount of data.

## Introduction

Health is basically a product of human lifestyle (1). The general recommendations for addressing diseases that are responsible for two-thirds of deaths worldwide are mainly related to lifestyle changes. Therefore, improper eating habits are the main cause of today's health problems. In a diet that is tailored to the individual's energy needs and interests, one can have a healthy body that is less susceptible to disease, which can have a major impact on one's long-term health. But since people are often very poor at judging their dietary health and need support to implement positive changes, long-term programs can be developed to help people make better decisions. However, this solution is neither practical nor economically justifiable for everyone.

In this regard, smart technologies offer a high potential to address these challenges, and among these, the recommender system not only provides recommendations that are in accordance with the user's preference, but can also consider the user's health (2). The recommender system suggests the most suitable items (data, information, goods, etc.) by analyzing the user's behavior. This system is an approach that is presented to face the problems caused by the large amount of information and helps the user to get closer to the goal faster among the huge amount of information.

By examining the choices of users in the past, these systems find patterns in the data and according to those behavioral patterns, show the appropriate recommendation for each user. Recommender systems are divided into 9 general categories: Collaborative Filtering models, Content-based recommender systems, Knowledge-based recommender systems, Demographic recommender systems, Group-based and Hybrid recommender systems, Text-based recommender systems, Time-sensitive recommender systems, Location-based recommender systems, and Social recommender systems (3).

Therefore, food recommender systems are recognized as a potential tool to help people eat better as part of a behavior change strategy that suggests better dietary changes, match to the user's needs, and more likely to be accepted and followed by the user (4).

Some food recommender systems combine these different aspects to create hybrid systems. Hybrid systems can combine the strengths of different types of recommender systems to create methods that can be used in a variety of settings (5).

In this paper, a hybrid recommender system is used, which includes collaborative filtering, content-based, and knowledge-based models. This system is designed by considering nutritional information and user preferences and food-health needs as a new research field to provide people with appropriate food selection recommendations based on food history. In this context, food recommender systems are also explored as a potential tool to help people eat healthier. It makes sense to use food recommender systems as part of a strategy to change the nutritional behavior of users. In this case, food recommender systems not only learn the user's preferences regarding food ingredients and styles, but also take into account their health problems, nutritional needs, and previous eating behaviors to suggest the healthiest food.

In addition, the use of new computational intelligence technologies, especially machine intelligence in the field of food recommender system, can be very important to accurately understand the user's food preferences in creating an effective food recommendation. Even for building a health-oriented food service, the user can only be encouraged to follow a recommendation if the recommended food matches the user's preferences. Thus, in this paper, learning algorithms such as Decision Tree, Bagging, AdaBoost, and $k$-Nearest Neighbors are used to create a food recommender system.

In summary, the innovations of the article are as follows:

- Providing a food recommender system according to the conditions and food menu in academic environments,

- Recommending food to students based on the characteristics of their energy needs and combining them with the history of their food choices,

- Creating an effective artificial intelligence model, especially the AdaBoost model in the field of food recommendation, considering the high dependence of eating habits and interest in different types of food.

## Background

In previous research, recommender systems in general and food recommender systems in particular have been investigated. Recommender systems were proposed in the mid-1990s, which automatically offer items of interest to each user according to preferences defined in the user's personal profile (6). Recommender systems use information filtering to recommend information of interest to the user and are defined as systems that recommend suitable products or services after learning the preferences and demands of customers (7, 8). Most recommender system research has focused on improving the accuracy of recommendation algorithms (9, 10). Specifically, they use background data, input data, and algorithms to make their recommendations (11). Therefore, the time needed by customers to search for items is saved (12), and they receive recommendations for suitable products to buy (13).

Moreover, these systems are particularly useful when they identify information previously unseen by the user (14). Recommender systems help users browse information on the web by making suggestions about them (15). They combine the ideas of information retrieval, user profiling, and machine learning to provide a more efficient search environment for the user (16). These systems are designed to learn user behavior to discover their preferences and either help them find what they are specifically looking for or what they might find useful in a vast amount of information (17). Recommender systems are utilized in various fields such as business, health, education, advertising, etc.

Furthermore, food recommender systems, focusing on parameters such as the type of recommended food, the source of the recommendation, and the location to get the recommended food, answer the questions that arise for users by making recommendations. While many recommender systems only try to give preferences to users in the fields of music, movies or books, recently these systems have also been used in the food domain to provide valid answers to the above questions. For example, Recipe key is a food recommender system that filters recipes based on favorite ingredients, food allergies, and item descriptions (such as types of foods, preparation time, etc.) selected by users. In relation to food consumption, lifestyle diseases such as diabetes and obesity, which are the cause of many chronic diseases, have increased these days (18). These consequences can be improved by using a proper diet. Food recommender systems generally differ from other recommender systems in many ways, such as considering the user's nutritional needs, weight goals, health problems, using food databases to recommend healthier foods, knowledge of nutrition, medicine and diet. Finally, food has unique characteristics, and in general, food recommender systems base their recommendations on user information, nutritional and health sources, and food recipes (19). Mika studies two types of food recommender systems (20). The first type recommends healthy recipes or foods that are most similar to recipes that the user has liked in the past. For example, it recommends recipes based on user search terms or ingredient input. The second type of recommender systems recommend only items to users that have already been identified by health care providers. In addition, there are two other types of food recommender systems (Type 3 and Type 4) that use other scenarios when recommending food. Considering both of the above criteria, the type 3 recommender system provides suggestions in order to balance the preferences and nutritional-health needs of the user. All these three types of recommender systems are primarily designed for specific users. However, type 4 represents group recommendations where the food is consumed by a group of users rather than by individuals.

## Methods

In this research, the food reservation information of the students collected from the website of the nutrition department of one of Iran's public universities has been used. The recommender system models based on machine learning methods are designed to suggest suitable food for students using modern nutrition sciences and according to the amount of energy required by the student and the student's food selection records.

In general, the research method is quantitative-qualitative. The qualitative part includes experts' opinions about the correctness of food energy content, and in the quantitative part, the method of recommender systems is used. The current food recommender system is an online system.

### Research Data

The data required for this research was collected by studying the students' nutrition reservation records and their profile questionnaire information in the university nutrition system and based on the recommendations of nutrition experts. These data are:

- **Profile and personal information of each student**

The data required to obtain the BMR index and the amount of calories needed by each student according to the student's daily activity and mobility have been received through the student health system in a standard and valid manner. In addition, students should choose their type of diet.

- **Data of each food**

According to the food recommendations in the recommender system of this research, it was necessary to determine the amount of calories and the percentage of food served in the university program for each food in the system. For this purpose, a modern science nutritionist was used to calculate food calories based on weight and ingredients. The obtained information was transferred to the system based on a new dataset.

### Modeling Food Recommender System for Students

The purpose of this modeling is to provide food recommendations according to the energy indicators required by each user and the history of the user's eating behavior. The type of meal, the type of diet, the difference between the calories needed by the student in each meal and the calories of the food, and the percentage of choosing each food in the past were considered as input variables. The features of the model are shown in Table 1. Group 1 for selected food and group 0 for non-selected food were considered as output items.

Each of the Decision Tree, *k*-NN, AdaBoost, and Bagging models were implemented with different combinations of parameters on the training data using Grid search, and when the best parameters were determined, they were evaluated on the test data. Accuracy, precision, recall and F1-score criteria were used to evaluate the models.

**Table 1:** Features of food recommender system model

| Description |
| --- |
| Lunch or dinner |
| Type of diet chosen by the student (1 to 4) |
| Difference between the calories needed by the student in the meal and the calories in the food |
| History of student meal reservation |

Python language was used to implement and create machine learning models. In general, first all models including Decision Tree, *k*-NN, AdaBoost, and Bagging were developed using training data. Then, the samples of the test set were used to validate their quality. To evaluate the models, the criteria of accuracy, precision, recall and F1-score were used.

- **Model evaluation method**

In this study, the efficiency evaluation method was implemented offline using the cross-validation approach with $k = 10$. Accuracy, precision, recall and F1-score were used to calculate efficiency. These criteria are the most famous and common criteria for measuring and evaluating the output of the recommender systems, which are explained below.

Precision and recall measures are used to evaluate the quality of the recommendations provided. Precision is the ratio of the number of correct recommendations given to the student to the total number of recommendations calculated from equation (1).

$$\text{Precision} = \frac{\text{Correct recommendations}}{\text{Total recommendations}} \tag{1}$$

Recall is the ratio of the number of recommendations provided for the student to the total number of recommendations available for the student and is calculated through equation (2).

$$\text{Recall} = \frac{\text{Correct recommendations}}{\text{Relevant recommendations}} \tag{2}$$

The F1-score criterion is called the harmonic mean of the two previously known criteria and is calculated based on equation (3).

$$\text{F1-score} = \frac{2 * \text{recall} * \text{Precision}}{\text{recall} + \text{Precision}} \tag{3}$$

- **Model output**

When choosing food by the student, according to the energy indicators needed by each person and the history of choosing among the foods presented in the desired meal, the best score obtained among the foods is recommended to the user.

## Research Process

The general research process is shown in Figure 1.

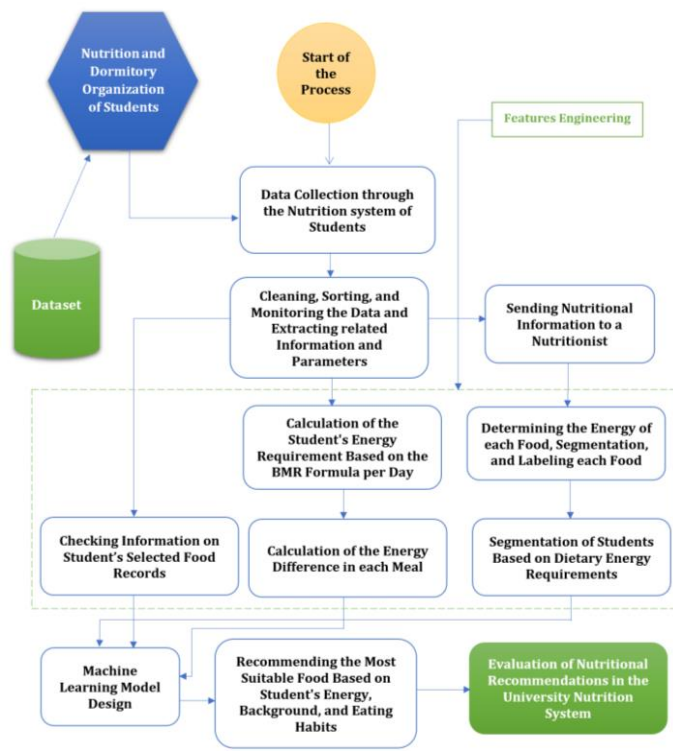

**Figure 1:** Research process

According to Figure 1, first, the data required for this research was collected by two methods: studying the students' meal reservation records and completing the personal information questionnaire in the university nutrition system, which includes the basic information of students in the university nutrition system, sports and dormitory activities system based on BMR indicators and the type of diet chosen by the student. Then, the standard calorie table for each food presented in the university menu based on its ingredients was prepared by the university nutritionist. Subsequently, in the pre-processing stage, since the data of this research is validated in the university software during registration, and invalid information is prevented by the system, there is no need to perform pre-processing techniques to prepare the data to increase the system efficiency. In order to extract the characteristics of the body mass index based on the student's height and weight, the basal metabolism was calculated based on gender, height, weight, age, activity level and the type of diet requested, and the food information in terms of calories that was completed by the nutritionist was automatically entered into the system. Then, Decision Tree, k-Nearest Neighbors, AdaBoost, and Bagging models were used. Furthermore, the difference between the calories consumed and the recommended calories according to the choice of diet (weight maintenance, weight loss, weight gain, health and traditional medicine) was included as the coefficient of the recommended meal in the data.

In the following, the data collected based on the meal is entered into the machine learning models under review, and after choosing the best machine learning method, the results are sent to the recommender function and the food system to suggest the most suitable food to the student when making a meal reservation based on the energy indicators and food selection history.

## Results

Research in the field of recommender systems is often based on the comparison of prediction accuracy, i.e. the better the evaluation scores, the better the recommender performs. However, it is difficult to compare the results of different recommender systems due to the many options in designing and implementing the evaluation strategy. In this section, the results obtained from the proposed learning models are examined. First, data description charts are presented, and then the obtained results are discussed.

### *Results of the Best Hyperparameters in Food Recommender System for Students*

To run and implement the models, it is first necessary to determine the best parameters of each model. For this purpose, the data was divided into two parts, training and test, with a ratio of 70 to 30. Then, four models including Decision Tree, k-Nearest Neighbors, AdaBoost, and Bagging were implemented for training data with 10-Fold and their hyperparameters were determined. In the following, all the models were implemented with their best parameters on the test data and finally evaluated with the criteria of accuracy, precision, recall and F1-score. Table 2 shows the best parameters of each model.

**Table 2:** Best parameters of the models in the food recommender system

| Model | Best Parameters |
|---|---|
| Decision Tree | Entropy criterion (maximum tree depth): 12<br>Minimum number of samples of an internal node: 3<br>Minimum number of samples required in one sheet: 3 |
| *k*-Nearest Neighbors | Number of neighbors: 13<br>Manhattan metric parameter and uniform weight |
| AdaBoost | SAMMER algorithm<br>Estimator *n*: 70 |
| Bagging | Maximum feature value: 0.7<br>Largest sample: 0.5<br>Estimator *n*: 20 |

### *Results of the Models Based on the Evaluation Criteria*

The output of Decision Tree, AdaBoost, KNN, and Bagging models according to the extracted features including precision, recall, F1-Score, accuracy, and AUC are shown in Table 3. According to the evaluation results, the best performance is obtained in the AdaBoost model and the worst performance in the KNN model.

**Table 3:** Results obtained based on the criteria evaluation (%)

| Model | Precision | Recall | F1-score | Accuracy | AUC |
|---|---|---|---|---|---|
| Decision Tree | 72.5 | 72.5 | 72.5 | 72.5 | 82 |
| AdaBoost | 73.5 | 73.5 | 74 | 73.7 | 83 |
| KNN | 69.5 | 69.5 | 70 | 69.5 | 77 |
| Bagging | 70 | 70 | 70 | 69.9 | 79 |

### *Comparison of Results and Evaluation Based on ROC Curve*

The area under the Receiver Operating Characteristic (ROC) curve, or AUC (area under curve) value is calculated to verify the reliability and evaluate the performance of the constructed diagnostic models. In this regard, the ROC curves of the four Decision Tree, Bagging, AdaBoost, and KNN models have been drawn. The closer the curves are to one, the more accurate the model is. The ROC curves for the four models are shown in Figure 2. According to the curves, it is clear that the ROC of the AdaBoost model is close to one. This means that this model has better and more accurate results than other models on this dataset.

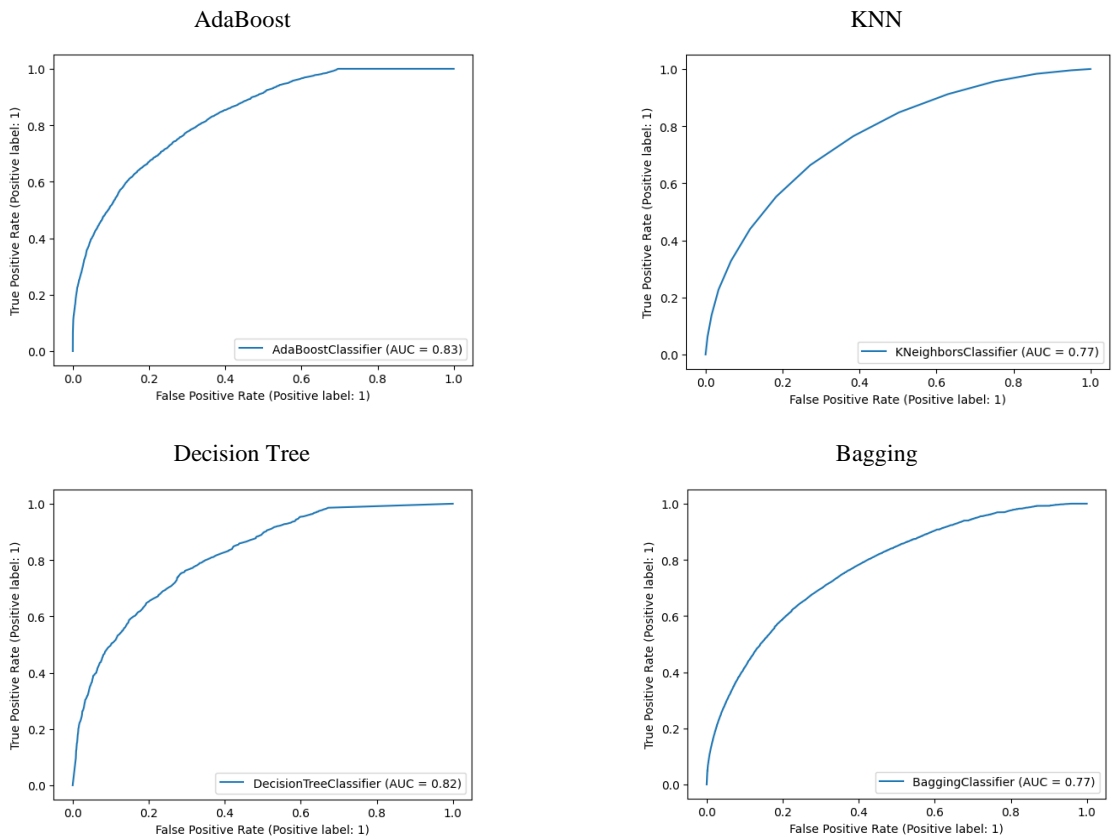

**Figure 2:** Receiver Operating Characteristic (ROC) curves of four models

## Proposed Food Recommender System

According to the evaluation results of the models, it can be claimed that the AdaBoost model has the best performance, when after developing the model, the results are sent to each user using the designed WebAPI. Figure 3 shows the result of recommending food to the user in the system. When choosing food by the student, considering two energy indicators and selection history, among the foods presented in the desired meal, the food with the highest score is recommended to the user.

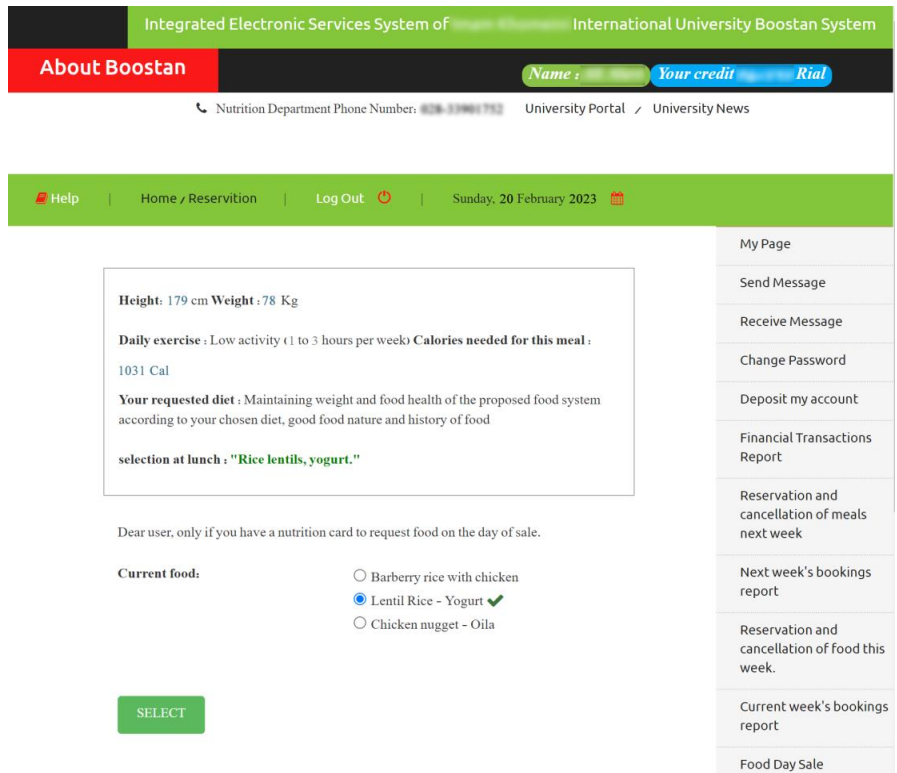

**Figure 3:** Example of recommending a student in the nutrition system using the food recommender system

## *Satisfaction and Performance Evaluation of Recommender System Recommendations*

In order to measure the level of satisfaction with the implementation of the food recommender system using the machine learning approach, two surveys was carried out regarding the recommendations made in the system and the level of satisfaction of the students. The first survey was conducted after 30 working days of providing recommendations to students. The second survey was conducted at the end of the research, which lasted for 90 days, and its results are shown in Table 4. In this research, out of 2519 students, 1637 participated in the system survey after receiving the recommendations.

**Table 4:** A report of the survey of students at the end of the research

| Questions | Level of Satisfaction in the Second Survey (%) | | | | |
|---|---|---|---|---|---|
| Are you satisfied with the implementation of the recommender system in the university? | Very Satisfied | Satisfied | No Opinion | Low Satisfied | Not Satisfied |
| | 39 | 36 | 13 | 8 | 4 |
| How satisfied are you with the recommendations of the system in your chosen diet? | Very Satisfied | Satisfied | No Opinion | Low Satisfied | Not Satisfied |
| | 43 | 36 | 11 | 7 | 3 |
| How satisfactory were the recommendations to achieve your desired diet? | Very Satisfied | Satisfied | No Opinion | Low Satisfied | Not Satisfied |
| | 42 | 33 | 16 | 7 | 2 |

## Discussion

Nowadays, many people skip a meal or turn to methods such as hydrotherapy and diuretics to lose weight, while such approaches cause physical harm and neither of them are helpful (21). The best way to control weight, whether in weight gain or loss diets, is to manage daily calorie intake, which the food recommender system suggests the best option by combining the amount of calories needed in each meal and the user's interests. Another application of this system is to provide group recommendations to the management of the university nutrition department based on the calories of each food and the amount of calories needed by each user (22).

In this system, first, the foods selected by the users are divided into segments according to their energy level; 5 segments are allocated to the amount of extra energy received in each meal and 5 segments are allocated to the amount of calories less than the users' needs. Each meal is reviewed in its own section. This analysis is performed over a 90-day time window frame and allows the nutrition management to decide and plan to increase or decrease the amount of food or the energy content of each meal based on the level of acceptance of the food and the energy required by the users. Figures 4 and 5 show the number of meals served based on the difference in energy required and received per meal. By checking these values, it is possible to reduce extra energy in foods or plan to reduce food waste.

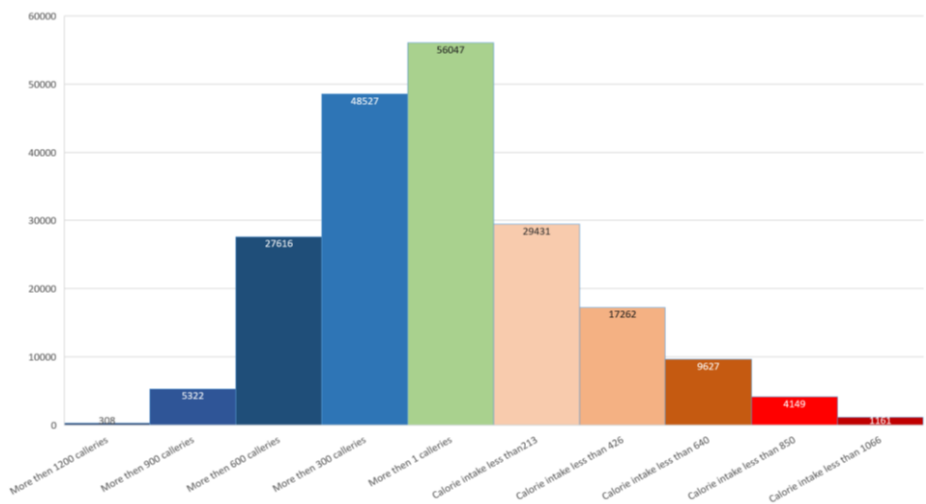

**Figure 4:** Graph of the difference in energy intake during lunch

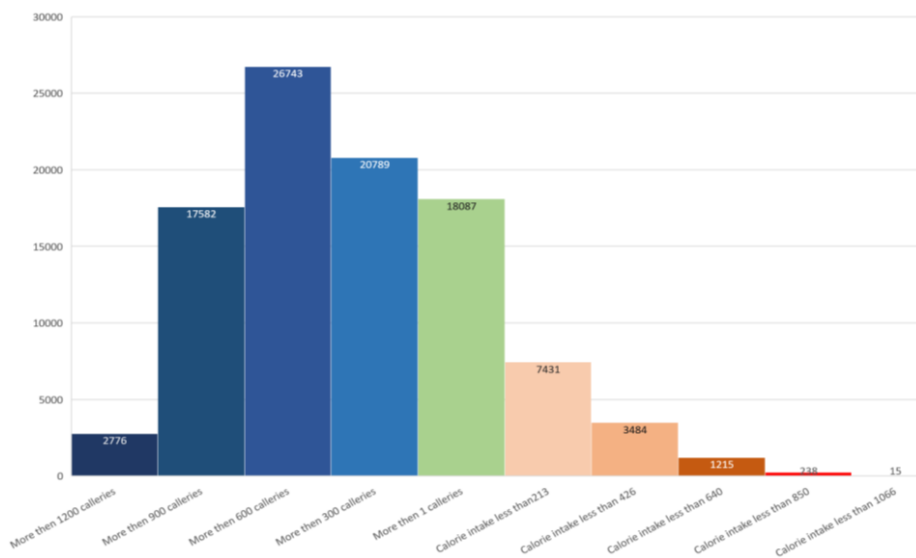

**Figure 5:** Graph of the difference in energy intake during dinner

# Conclusion

Today, despite the ever-increasing variety and abundance of foods and due to the lack of attention to healthy diet, the possibility of suffering from dietary diseases such as diabetes is increasing. On the other hand, people's attention to collecting information on diet and obtaining information through the Internet has expanded in order to prevent the development or prevalence of diseases. But due to the scattering of information, they do not reach the desired result. To deal with these issues, a new field has emerged to obtain information of interest among a huge amount of related and unrelated data, which is known as recommender systems. The study and presentation of the food recommender system for students in the university environment with an approach to adjust the university menu has been very few. On the other hand, in the present study, to recommend food to the student, the data of the student's food reservation (food history) and the amount of calories required in each meal were examined in order to make the food recommendation according to the student's requested diet. Therefore, in this paper, considering the importance of lifestyle, artificial intelligence learning models such as Decision Tree, Nearest Neighbors, Adaboost and Begging were designed and developed in the field of food recommender systems. These systems use personal health information as information sources. Based on the evaluation results of the models, the AdaBoost model had the best performance in terms of accuracy with a rate of 73.70%. As a future work, an intelligent system can be designed to recommend the suitable food for the student by creating a model based on a combination of the student's temperament, traditional medicine and taste, food records and the amount of energy required. The use of deep learning methods to increase and improve the accuracy of the models, providing additional recommendations regarding the foods received by students in the university, such as required snacks, and examining the effect of students' diet on the quality of their learning can be the subject of future studies.